\documentclass[aps, pre, eqsecnum, amsmath,amssymb,showpacs]{revtex4}
 \usepackage[final]{graphicx}  

\newcommand{\dd}{\text{d}}
\newcommand{\oD}{\overline{D}}
\newcommand{\om}{\overline{m}}
\newcommand{\oalpha}{\overline{\alpha}}
\newcommand{\oepsilon}{\overline{\epsilon}}
\newcommand{\ophi}{\overline{\phi}}

\begin{document}
\renewcommand{\thefootnote}{\fnsymbol{footnote}}

\title{Magnetization of concentrated polydisperse ferrofluids: Cluster
expansion}
\author{B.~Huke and M.~L\"{u}cke \\}
\affiliation{Institut f\"{u}r Theoretische Physik,Universit\"{a}t des Saarlandes,
D-66041 Saarbr\"{u}cken, Germany \\}

\renewcommand{\thefootnote}{\arabic{footnote}}
\setcounter{footnote}{0}
\date{\today}


\begin{abstract}

The equilibrium magnetization of concentrated ferrofluids
described by a system of {\it polydisperse} dipolar hard spheres is calculated 
as a function of the internal magnetic field using the
Born--Mayer or cluster expansion technique. This paper extends the results
of Phys. Rev. E {\bf 62}, 6875 (2000) obtained for {\it monodisperse\/} ferrofluids.
The magnetization is given as a power 
series expansion in two parameters related to the volume 
fraction and the coupling strength of the dipolar interaction, respectively.

\end{abstract}

\maketitle


\section{Introduction}
\label{intro}

Ferrofluids \cite{R85} are suspensions of ferromagnetic particles of about 
10 nm diameter in a carrier fluid. The particles are stabilized against 
aggregation by coating with polymers or by electrostatic repulsion of charges 
brought on their surface. As long as the concentration of the particles is
low, the equilibrium magnetization of a ferrofluid is that of an ideal 
paramagnetic gas. In highly concentrated ferrofluids on the other hand, the
magnetization is influenced by effects of particle--particle interactions.

We studied these effects for ferrofluids that are described by a system of 
identical
dipolar hard spheres in \cite{HL}, from now on referred to as paper~I.
In that paper we used the technique of the Born--Mayer or cluster expansion
technique to evaluate the equilibrium magnetization as
a series expansion in terms of the volume fraction $\phi = N \pi D^3/6V$, and
a dipolar coupling parameter $\epsilon = m^2/4 \pi \mu_0 kT$, with $N/V$
being the particle density, and $D$ and $m$ being the common hard sphere
diameter and magnetic moment of the particles, respectively.

However, real ferrofluids are polydisperse, i.~e.\ the particles vary in size and 
magnetic moment. This property has a strong influence on the equilibrium 
magnetization, for concentrated as well as for dilute fluids. The goal of
this paper is to generalize the findings of paper~I to include the effects
of polydispersity. 

The linear response problem of determining the static initial susceptibility
 $\chi$ of a mixture of dipolar hard spheres was investigated already for the 
equivalent electric case in the framework of integral theories: 
The mean spherical model \cite{W71} was extended to binary or multicomponent 
mixtures \cite{AD73,IB74,HS78,FHI79,RH81,CB86}. The  
reference hypernetted chain method \cite{FP84} was also applied to bidisperse 
systems \cite{LL87,LL89}. Recently \cite{KS02}, the mean spherical model was 
used within the algebraic perturbation theory \cite{K99}, however without
leading to new results for the initial susceptibility. The mean spherical model
was also extended to polydisperse ferrofluids in arbitrary high fields
\cite{SPMS90,ML90}. Another theory dealing with arbitrary fields is the
high temperature approximation \cite{BI92}. A variant of this theory was
proposed in \cite{PML96} and extended in \cite{IK01}.

Our calculation follows closely that of paper~I. Therein the application of 
the cluster expansion technique to a monodisperse system of dipolar hard 
spheres resulted in an expression for the magnetization $M$ that can
be put into the form 
\begin{equation}
M = M_{\text{sat}} \sum_{m=0}^\infty \sum_{n=0}^\infty \phi^m \epsilon^n
L_{m,n} (\alpha) \;\; , 
\end{equation}
where $M_{\text{sat}}$ is the saturation magnetization of the fluid. 
The functions $L_{m,n} (\alpha)$ were given explicitly in terms of analytic 
expressions in the dimensionless 
magnetic field $\alpha$. We calculated $L_{2,2}(\alpha)$ and some 
of the $L_{1,n}(\alpha)$. Lower orders vanish, except for the Langevin function 
$L_{0,0}(\alpha)$. 

In the polydisperse case discussed here the parameters $\phi$, $\epsilon$ and 
$\alpha$ are replaced by more
generally defined quantities $\ophi$, $\oepsilon$ and $\oalpha$ 
(cf. Sec.~\ref{SEC3}). The 
calculated $L_{m,n}$ transform into one--, two-- or threefold sums over all 
particles, where the individual addends are analytical functions of the 
magnetic moments and diameters of the involved particles, and the reduced 
magnetic field $\oalpha$.

The paper is organized as follows. In Sec.~\ref{SEC2} we explain the
principles of the cluster expansion technique. The main part of the paper is 
Sec.~\ref{SEC3}, where we generalize the results of paper~I for the 
equilibrium magnetization in the monodisperse case to polydisperse 
ferrofluids. The findings are discussed in Sec.~\ref{SEC4} using example
distributions. 
In Sec.~\ref{SEC4a} the results are compared to experimental data.
We conclude in Sec.~\ref{SEC5}.

\section{Cluster expansion: Application to the system of dipolar hard
spheres}
\label{SEC2}

Here we recapitulate briefly the principle of the Born--Mayer or cluster 
expansion technique: Consider a system of particles $i = 1, \, ...\, N$ 
interacting with an 
external potential $V_i$ and with each other via a potential $V_{ij}$. 
To calculate thermodynamic properties of the system one has to find the
canonical partition function
\begin{equation}
Z = \int e^{ -\sum_k v_k - \sum_{i<j} v_{ij}} \;
d \Gamma \;\; . \label{candistrigen}
\end{equation} 
Here $v_i = V_i/kT$, $v_{ij} = V_{ij}/kT$, and $d \Gamma$ means integration 
over the configuration space. The kinetic energy of the particles, if important,
can be thought to be included in the terms $V_i$. One now writes
\begin{equation}
Z = \int \prod_k e^{-v_k}
\prod_{i<j} (1 + f_{ij} ) \; d \Gamma \;\; ,
\label{Zinf}
\end{equation} 
where
\begin{equation}
f_{ij} = e^{- v_{ij}} -1 \;\; .
\end{equation} 
If the typical interaction energy is small compared to $kT$, the $f_{ij}$ can 
be considered as small parameters for the expansion of the integrand in 
Eq.~(\ref{Zinf}).
The leading terms factorize into low dimensional integrals that can be 
calculated at least numerically.

In the system of dipolar hard spheres (monodisperse or polydisperse) the 
interaction potential $V_{ij}$ consists of a dipole--dipole (DD) interaction 
and a hard core (HC) repulsion part,
$V_{ij} = V^{DD}_{ij} + V^{HC}_{ij}$, where the first part is given by
\begin{equation}
V^{DD}_{ij} = 
- \frac{ 3 ({\bf m}_i \cdot {\bf \hat{r}}_{ij} )({\bf m}_j \cdot 
{\bf \hat{r}}_{ij} ) - {\bf m}_i \cdot {\bf m}_j }
{4 \pi \mu_0 r_{ij}^3} \;\; , \label{VDDijdef}
\end{equation}
for two particles with magnetic moments ${\bf m}_i$ and ${\bf m}_j$ at a
distance ${\bf r}_{ij} = {\bf x_i} - {\bf x_j}$,  with 
$r_{ij} = |{\bf r}_{ij}|$, and ${\bf \hat{r}}_{ij} = {\bf r}_{ij}/r_{ij}$.
For particles with diameters $D_i$ and $D_j$ one has $V_{ij}^{HC}(r_{ij}) = \infty$,
if $r_{ij} < D_{ij} = (D_i+D_j)/2$, and $V_{ij}^{HC}(r_{ij}) = 0$ otherwise.

Taking the thermodynamic limit in a system of dipolar particles requires some
care because of the long range character of the forces \cite{BGW98}. 
We circumvented this problem by decomposing the dipolar potentials into a short
range and a long range part, and replacing the latter by an effective mean
field. Within this approach a particle experiences the local magnetic field
\begin{equation}
{\bf H}_{local} = {\bf H}_s + {\bf H}_{dipole,near} = 
{\bf H} + \frac{{\bf M}}{3} + {\bf H}_{dipole,near} \;\; .
\end{equation}
It consists of the dipolar near field ${\bf H}_{dipole,near}$ that is produced 
by the other particles within a sphere of radius $R_s$ and of an effective 
"external" field
\begin{equation} 
 {\bf H}_s = {\bf H} + \frac{\bf M}{3} \;\; , \label{Hsdef}
\end{equation}
seen by the particle in question at the center of the sphere. Here ${\bf H}$ is
the macroscopic internal magnetic field and ${\bf M}$ the sought after 
equilibrium magnetization. Thus, when
evaluating the partition function one has to take 
\begin{equation}
V_i = - {\bf m}_i \cdot {\bf H}_s 
\end{equation} 
as the external potential. The radius $R_s$ of the sphere has to be taken to be
sufficiently large to allow the far-field dipolar contributions to be replaced by
those of a continuum -- cf. paper~I for details. Neither the kinetic energy
of the magnetic particles, nor the carrier fluid has to be taken into account in 
the partition function, since these terms do not contribute
to the equilibrium magnetization. The configuration space is thus given by
the positions ${\bf x_i}$ of all particles and the orientations $\Omega_i$
of their magnetic moments: $d \Gamma = d^n {\bf x}_i d^n \Omega_i$. 

In paper~I we used in addition also an expansion in the dipolar interaction: 
The $f$--terms were expanded as
\begin{equation}
f_{ij} = e^{-v^{HC}_{ij}} e^{-v^{DD}_{ij}} -1 
= f_{ij}^{(0)} + f_{ij}^{(1)} + f_{ij}^{(2)} + ... ,
\end{equation}
with
\begin{eqnarray}  
f_{ij}^{(0)} &=& e^{-v^{HC}_{ij}} - 1\\
f_{ij}^{(n)} &=& \frac{\left(-v^{DD}_{ij}\right)^n}{n!} 
e^{-v^{HC}_{ij}} \label{fijndef}
... \;\; ,  n \geq 1 \;\;.
\end{eqnarray}
The two expansions concerning the $f$--terms and $v_{ij}^{DD}$ together 
translate in the monodisperse case into a double power expansion of $Z$ in the 
the volume fraction $\phi$ of the particles and the dipolar coupling constant 
$\epsilon$. We calculated the terms in 
$O(\phi \epsilon^n)$ and in $O(\phi^2 \epsilon^2)$ of $Z$ and from that the 
equilibrium magnetization in the same order.

\section{Calculating the equilibrium magnetization}
\label{SEC3}

\subsection{Notation}
\label{SEC31}

Consider a system of $N$ spherical hard particles with diameters 
$D_1$, ..., $D_N$ carrying permanent magnetic moments $m_1$, ..., $m_N$
contained in a volume $V$ and subjected to a magnetic field ${\bf H}$. 
Let $\oD$ and $\om$ be some "typical" values for diameters and magnetic 
moments that are discussed further below. We then define the parameter 
$\ophi$ related to the volume fraction
of the hard spheres and the dipolar coupling parameter $\oepsilon$ as 
\begin{equation}
\ophi = \frac{N\pi \oD^3}{6 V} \;\; ,\qquad
\oepsilon = \frac{\om^2}{4 \pi \mu_0 kT \oD^3} \;\; .
\end{equation}
The equilibrium magnetization is calculated as a power expansion in these 
two parameters. The dimensionless magnetic fields are defined as
\begin{equation}
\oalpha = \frac{\om H}{k T} \;\; , \;\; \oalpha_s = \frac{\om H_s}{k T} \;\; .
\end{equation}
Diameters and magnetic moments will be expressed in units of the typical
values via
\begin{equation}
\Delta_i = \frac{D_i}{\oD} \;\; , \qquad \mu_i = \frac{m_i}{\om} \;\; .
\end{equation}
We will also use the minimal possible distance between two hard spheres 
$i$ and $j$ given by
\begin{equation}
\Delta_{ij} = \frac{1}{2} \left( \Delta_i + \Delta_j \right) = 
\frac{D_{ij}}{\oD} \;\; .
\end{equation}
Furthermore we introduce the reduced magnetic fields for each particle $i$ by
\begin{equation}
\alpha_i = \frac{m_i H}{k T} = \mu_i \oalpha \;\; ,\;\;
\alpha_{si} = \frac{m_i H_s}{k T} = \mu_i \oalpha_{s} \;\; .
\end{equation}

Our cluster expansion does not depend on how the characteristic 
values of $\oD$ and $\om$ are defined in detail. For example they could be taken
as as some weighted mean of the $D_i$ and $m_i$, respectively, or their most
probable values. To preserve this 
freedom of choice in our expansion offers some advantages for the comparison 
with magneto--granulometric analyses where the distribution of the diameters
and magnetic moments is not known a priori but on the contrary the goal of the 
calculations. 

Note, however, that $\ophi$ coincides with the actual 
volume fraction $\phi$ of the hard spheres only if one defines $\oD$ via the 
mean volume of the particles
\begin{equation}
\oD^3 = \frac{1}{N} \sum_i D_i^3
= \int D^3 P(D) \, dD \equiv \left< D^3 \right>_P \;\; . 
\label{defDbar}
\end{equation}
Here $P(D)$ is the normalized distribution function of the hard sphere
diameters. 

Similarly $\om$ is related to the saturation magnetization 
$M_{\text{sat}}$ of the ferrofluid via $M_{\text{sat}} = N \om/\mu_0 V$ only if
$\om$ is defined by 
\begin{equation}
\om = \frac{1}{N} \sum_i m_i = \int m(D) P(D) \, dD \;\; . 
\label{defmbar}
\end{equation}
The second equality of Eq.~(\ref{defmbar}) holds when the magnetization of each
particle is given by a function of its volume. We assume that this is the case
and thus describe in this paper polydispersity effects of the ferrofluid by a 
distribution function $P(D)$ depending only on the hard sphere diameter $D$.
The generalization to a distribution 
function $P(D,m)$ of independently varying diameters and moments is 
straightforward. Averages weighted with the distribution function $P(D)$ of the
diameters will mostly appear in the reduced version as integrals over the
reduced diameter $\Delta = D/\oD$ with the appropriate weight function 
$P(\Delta)$.

The thermodynamic mean with respect to the noninteracting system will be
denoted by $\left< ... \right>_0$ and the corresponding canonical
partition function by $Z_0$. With this notation integrals over the $f$--terms
appearing in $Z$ (\ref{Zinf}) can be written in the form
\begin{equation}
\int \prod_k e^{-v_k} f ...  \; d \Gamma = Z_0 \left< f ... \right>_0 \;\; .
\end{equation}
But in contrast to paper~I we derive here an approximation directly for the 
free energy $F = - kT \ln Z$. If the particle--particle
interaction would depend only on interparticle distance then $F$ would be given
by 
\begin{equation}
F = F_0 - 
k T \left( \frac{1}{2} {\sum}^\prime \left< f_{ij} \right>_0 + 
\frac{1}{6} {\sum}^\prime \left< f_{ij} f_{jk} f_{ki} \right>_0 \right) \;\; ,
\label{Fseries}
\end{equation}
including orders up to $O(\ophi^2)$, or, more generally speaking, up to terms 
of second order in the number density. The primed sums are taken over all
particle pairs $i$,$j$, resp.\ all triples $i$,$j$,$k$.
While Eq.~(\ref{Fseries}) does not hold for a system of dipolar particles in a 
magnetic field in arbitrary order of $\oepsilon$ it still is correct in the 
orders we want to calculate.
 
The polydisperse generalization affects the calculation of the integrals
in Eq.~(\ref{Fseries}) in two ways: ({\it i}) The fact that the individual dimensionless
magnetic fields $\alpha_i$ are different leads to more complicated 
expressions for some resulting functions compared to the monodisperse case
-- see the definitions of $G^P_n$ and $K^P$ below. ({\it ii}) 
The dispersion in the hard sphere diameters requires more difficult geometrical
considerations concerning the $v_{ij}^{HC}$--terms, especially in the three
particle integral.

\subsection{The leading term: polydisperse Weiss model}
\label{SEC32}

The leading term in Z is the partition function of the (formally) 
noninteracting paramagnetic gas in the magnetic field $H_s$
\begin{equation}
Z_{0} = \int \prod_k e^{-v_k} \; d \Gamma = \prod_k z_k \;\; ,
\end{equation}  
\begin{equation}
z_k = \int e^{-v_k} d {\bf x}_k d \Omega_k = 
4 \pi V \frac{\sinh \alpha_{sk}}{\alpha_{sk}} \;\; .
\end{equation}  
The equilibrium magnetization $M(\oalpha_s)$ obtained from
\begin{equation}
M(\oalpha_s) = -\frac{1}{\mu_0 V} \frac{\partial F}{\partial H_s} = 
\frac{1}{\mu_0 V} \frac{\partial (k T \ln Z)}{\partial H_s} = 
\frac{\om}{\mu_0 V} \frac{\partial \ln Z}{\partial \oalpha_s} 
\end{equation} 
reads in leading order 
\begin{equation}
M(\oalpha_s)= \frac{N \om}{\mu_0 V} {\cal L}_{\text{poly}}(\oalpha_s) \;\; . 
\label{Mofalphas}
\end{equation}
Here 
\begin{equation}
{\cal L}_{\text{poly}}(\oalpha_s) = \frac{1}{N} \sum_i \mu_i 
{\cal L} \left( \mu_i \oalpha_s \right)  
= \int \mu(\Delta)
{\cal L} \left[ \mu(\Delta)  \oalpha_s \right] P(\Delta) d \Delta \;\; .
\label{Lpoly}
\end{equation}
is given by the sum of the Langevin paramagnetic contributions coming from
each (reduced) magnetic moment $\mu_i=m_i/\om$ with ${\cal L}$ being the 
Langevin function. The second equality in Eq.~(\ref{Lpoly}) is the continuous
analog of the sum with $\mu(\Delta)=m(\Delta)/\om$ and $\Delta=D/\oD$.
If one defines $\om$ via $N \om = \sum_i m_i$ so that 
$N \om/\mu_0 V = M_{\text{sat}}$ then 
${\cal L}_{\text{poly}}(\oalpha_s \rightarrow \infty) = 1$.

The result (\ref{Mofalphas}) reduces to the well known expression for the
magnetization of a polydisperse ideal paramagnetic gas as a superposition of
Langevin functions, if one replaces $\oalpha_s$ by $\oalpha$ 
(see e.~g.~\cite{ML90}). However, the dipolar far-field contributions enter 
via (\ref{Hsdef}) as a mean field into 
\begin{equation}
\oalpha_s = \oalpha + \om M/3kT \;\; .
\label{oalphas}
\end{equation} 
Thus, the lowest order result (\ref{Mofalphas}) for $M(\oalpha)$  
\begin{equation}
M = \frac{N \om}{\mu_0 V} {\cal L}_{\text{poly}} \left( \oalpha + 
\frac{\om M}{3kT} \right) 
\label{Mofalpha+M}
\end{equation}
contains already corrections from the particle--particle interaction
in the mean field approximation and (\ref{Mofalpha+M}) is the polydisperse 
generalization of the Weiss model \cite{Ce82}. 

By replacing M on the right
hand side of Eq.~(\ref{Mofalpha+M}) by the expression for the ideal paramagnetic 
gas, one arrives at the equation proposed in \cite{PML96}.

\subsection{The magnetization in $O(\ophi)$ }
\label{SEC33}

To calculate the canonical partition function in linear order of $\ophi$
we follow the lines of paper~I, Sec.~IV.
One needs to include only the linear $f$--terms in the expansion 
(\ref{Fseries}). Thus we write
\begin{equation}
F = F_0 - k T \frac{1}{2} {\sum}^\prime \left< f_{ij} \right>_0 \;\; .
\label{FexpinOphi}
\end{equation} 

For the second term in (\ref{FexpinOphi}), the trivial integrations over the 
degrees of freedom of all particles except $i$ and $j$ are performed 
first. This gives
\begin{eqnarray}
&& \frac{1}{2} {\sum}^\prime \left< f_{ij} \right>_0 = 
\frac{1}{2 Z_0} {\sum}^\prime \int \prod_k e^{-v_k} f_{ij}  \; d \Gamma 
\\ && \qquad
={\sum}^\prime \frac{1}{2 z_i z_j} \int e^{-v_i - v_j} f_{ij} 
\; d{\bf x}_i d{\bf x}_j d\Omega_i d\Omega_j \;\; . \nonumber
\end{eqnarray} 
Now we expand $f_{ij}$. Let 
\begin{equation}
A_{n,ij} = \int e^{-v_i - v_j} f^{(n)}_{ij} 
\; d{\bf x}_i d{\bf x}_j d\Omega_i d\Omega_j \;\; ,
\end{equation} 
such that 
\begin{equation}
F = F_0 - {\sum}^\prime \frac{kT}{2 z_i z_j} \sum_{n=0}^\infty A_{n,ij} \;\; .
\label{FexpinOphiwithA}
\end{equation} 

We need not calculate $A_0$, since this term does not contribute to the equilibrium
magnetization. Furthermore $A_1 = 0$, because a dipolar magnetic field vanishes when 
averaged over a spherical surface. This is explained in more detail in paper~I. 
Using the definition (\ref{fijndef}) of $f^{(n)}_{ij}$, and the dipolar 
potential (\ref{VDDijdef}), we can write
\begin{eqnarray}
A_n &=& \frac{V}{n!} \int 
e^{\alpha_{si} \cos \vartheta_i + \alpha_{sj} \cos \vartheta_j} \nonumber \\
&& \times \left( \frac{\om^2 \mu_i \mu_j}{4 \pi \mu_0 k T r^3_{ij}} \right)^n 
P^n(\varphi_i,\vartheta_i,\varphi_j,\vartheta_j,\varphi,\vartheta)
\label{Anresult1}\\
&& \times e^{-v^{HC}_{12}}
r^2_{ij} dr_{ij}  d\omega_{ij} d\Omega_i  d\Omega_j \;\; . \nonumber
\end{eqnarray}
Here we have integrated over ${\bf x_i}$ and decomposed ${\bf r}_{ij}$ into
the distance $r_{ij}$ and a spherical angle $\omega_{ij}$. The angles
$(\varphi_i,\vartheta_i)$, $(\varphi_j,\vartheta_j)$, and 
$(\varphi,\vartheta)$ represent the spherical angles $\Omega_i$, $\Omega_j$,
and $\omega_{ij}$, respectively. The function
\begin{equation}
P(\varphi_i,\vartheta_i,\varphi_j,\vartheta_j,\varphi,\vartheta) = 
3 (\hat{\bf m}_i \cdot {\bf \hat{r}}_{ij} )(\hat{\bf m}_j \cdot 
{\bf \hat{r}}_{ij} ) - \hat{\bf m}_i \cdot \hat{\bf m}_j 
\end{equation}
comes from the dipolar interaction.
The integration over the directions of ${\bf r}_{ij}$, ${\bf m}_i$, and
${\bf m}_j$ can still be done analytically. But in contrast to the monodisperse
calculation, the result is now a function of two parameters $\alpha_{si}$
and $\alpha_{sj}$. We define
\begin{eqnarray} \label{Gndef}
&&G^P_n(\alpha_{si},\alpha_{sj}) = \frac{V^2}{n! (n-1) \pi z_i z_j} 
\nonumber \\ 
&& \;\;\; \times 
\int e^{\alpha_{si} \cos \vartheta_i + \alpha_{sj} \cos \vartheta_j} \\
&& \qquad \times  
P^n(\varphi_i,\vartheta_j,\varphi_i,\vartheta_j,\varphi,\vartheta) 
\; d\Omega_i d\Omega_j d\omega_{ij} \;\; . \nonumber
\end{eqnarray}
$G^P_n(\alpha_{si},\alpha_{sj})$ is symmetric in its two arguments and a
polydisperse counterpart to the function $G_n(\alpha_{s})$ defined in 
paper~I. It is $G^P_n(\alpha_{s},\alpha_{s}) = G_n(\alpha_{s})$.
Some of the $G_n^P$ are given in the appendix.

Inserting Eq.~(\ref{Gndef}) into Eq.~(\ref{Anresult1}), integrating over 
$r_{ij}$ between the minimal distance $D_{ij}$ and $\infty$, and 
introducing $\ophi$ and $\oepsilon$ yields
\begin{eqnarray}
A_n &=& \frac{2}{N} z_i z_j\ophi \oepsilon^n (\mu_i \mu_j)^n 
\Delta_{ij}^{3-3n} G^P_n(\alpha_{si},\alpha_{sj}) \;\; , 
\end{eqnarray}
and together with (\ref{FexpinOphiwithA}) the free energy
\begin{eqnarray}
F &=& F_0 + ... \\ 
&&  - \frac{k T}{N} \sum_{n=2}^\infty \ophi \oepsilon^n 
{\sum}^\prime  (\mu_i \mu_j)^n 
\Delta_{ij}^{3-3n} G^P_n(\alpha_{si},\alpha_{sj})  \;\; . \nonumber 
\end{eqnarray}
Here the dots 
represent the contribution from $A_0$ that was not calculated. It can easily 
be shown that $F$ does not depend on a particular definition of $\om$ or
$\oD$.

Now, the equilibrium magnetization $M(\oalpha_s)$ is given in $O(\ophi)$ by
\begin{equation}
M(\oalpha_s)= 
\frac{N \om}{\mu_0 V}  \left[ {\cal L}_{\text{poly}}(\oalpha_s) + 
\sum_{n=2}^\infty  
\ophi \oepsilon^n G_{\text{poly},n}^{\prime}(\oalpha_s) \right] \;\; . 
\label{MofalphasOphi}
\end{equation}
The function $G_{\text{poly},n}^{\prime}$  is the derivative of 
\begin{eqnarray}\label{Gpolydef}
G_{\text{poly},n}(\oalpha_s) &=& \frac{1}{N^2} {\sum}^\prime 
(\mu_i \mu_j)^n \Delta_{ij}^{3-3n} 
G^P_n(\alpha_{si},\alpha_{sj}) \nonumber \\ 
&=& \int [\mu(\Delta_i) \mu(\Delta_j)]^n 
\Delta_{ij}^{3-3n} \nonumber \\
&\times& G^P_n(\alpha_{si},\alpha_{sj}) P(\Delta_i) P(\Delta_j) d \Delta_i d \Delta_j 
\end{eqnarray}
which is is a generalization of $G_n$ \cite{HL} and reduces to the latter in 
the monodisperse case $D_i = \oD = D$ and $m_i = \om = m$. 

In a last step, we convert the expression for $M$ as a 
function of $\oalpha_s$ into a function of $\oalpha$ using the definition 
of $H_s$ in Eq.~(\ref{Hsdef}). By expanding and iterating in a way that is
analogous to the procedure in paper~I, Sec.~IV~C we obtain the 
final result for $M$ up to order $\ophi$ 
\begin{eqnarray}
M(\oalpha) &=& \frac{N \om}{\mu_0 V } \Big[ {\cal L}_{\text{poly}}(\oalpha) +
8 \ophi \oepsilon {\cal L}_{\text{poly}}(\oalpha)
{\cal L}^\prime_{\text{poly}}(\oalpha)  \nonumber \\
&& \qquad \qquad+ 
\sum_{n=2}^\infty  \ophi \oepsilon^n G_{\text{poly},n}^\prime(\oalpha)
\Big] \;\; . \label{Mofalpha}
\end{eqnarray}
The two leading terms can be seen as the polydisperse extension of the high 
temperature approximation derived in \cite{BI92} for monodisperse systems.

\subsection{The contribution in $O(\ophi^2 \oepsilon^2)$}
\label{SEC34}

For the monodisperse system, the magnetization contribution in 
$O(\ophi^2 \oepsilon^2)$ was calculated in Sec.~V and appendix~B 
of paper~I. The cluster
integrals needed in that order are shown in Fig.~3 of paper~I. Some of them
vanish for the same reason as the contribution $A_1$ in $O(\ophi \oepsilon)$:
They involve the averaging of a dipolar field over a spherical surface.
Most of the remaining integrals cancel when the free energy $F = - kT \ln Z$
is calculated. Up to $O(\oepsilon^2)$ we can write the remaining term in
Eq.~(\ref{Fseries}) as
\begin{eqnarray}
\left< f_{ij} f_{ik} f_{jk} \right>_0 &=& 
\left< f^{(0)}_{ij} f^{(0)}_{ik} f^{(0)}_{jk} \right>_0 +
3 \left< f^{(2)}_{ij} f^{(0)}_{ik} f^{(0)}_{jk} \right>_0 \nonumber \\
&& +
3 \left< f^{(1)}_{ij} f^{(1)}_{ik} f^{(0)}_{jk} \right>_0 \;\; .
\label{FtermsinOphi2}
\end{eqnarray}
These three terms correspond to the graphs E, G and H in paper~I. The first one 
is an $O(\ophi^2 \oepsilon^0)$--term that does not contribute to the
magnetization. 

\subsubsection{Graph G}

Here we give an outline of the polydisperse generalization of the
calculation pertaining to graph G (appendix~B.7 of paper~I). The quantity 
$Z_G$ calculated in paper~I, is given by the sum
\begin{equation}
Z_G = Z_0 \frac{1}{2} {\sum}^\prime 
\left< f^{(2)}_{ij} f^{(0)}_{jk} f^{(0)}_{ki} \right>_0 
\end{equation}
over distinct particles $i, j, k$.
For calculating $\left< f^{(2)}_{ij} f^{(0)}_{jk} f^{(0)}_{ki} \right>_0$ one
starts with the trivial integrations: the degrees of freedom of all particles
except $i$, $j$, and $k$, the position of the center of mass of the three
remaining particles, and the orientation of ${\bf m}_k$, since particle $k$
is not involved in dipolar interactions in this cluster. Then, for a fixed
distance $r_{ij}$ the integrations over $\omega_{ij}$, $\Omega_i$, and
$\Omega_j$, defined as in Sec.~\ref{SEC33} are carried out. This introduces
the function $G_2^P$ into the result:
\begin{eqnarray}
&& \left< f_{ij}^{(2)} f_{ik}^{(0)} f_{jk}^{(0)} \right>_0 = 
\frac{36}{\pi N^2} (\mu_i \mu_j)^2 \ophi^2 \oepsilon^2 
G^P_2(\alpha_{si},\alpha_{sj}) \nonumber \\
&& \qquad \times \int
r_{ij}^{-4} e^{-v_{ij}^{HC}}  
 f_{ik}^{(0)} f_{jk}^{(0)} \; 
d r_{ij} d {\bf r}_{ik} \;\; .\label{GGfirst} 
\end{eqnarray}
The integral over ${\bf r}_{ik}$ can be described by the following geometrical 
considerations:
The volume of possible positions of particle $k$ has to be found, such that
this particle overlaps with both, particle $i$ (i.~e.\ $r_{ik} < D_{ik}$)
and $j$ ($r_{jk} < D_{jk}$). Otherwise the integral would vanish
because of the factor $f_{ik}^{(0)} f_{jk}^{(0)}$. This is only possible, if 
$r_{ij} < D_{ik} + D_{jk}$. 

In a final step the integration over $r_{ij}$
between $D_{ij}$ and  $D_{ik} + D_{jk}$ is carried out. The final result is
\begin{eqnarray}
\left< f_{ij}^{(2)} f_{ik}^{(0)} f_{jk}^{(0)} \right>_0 &=&
\frac{3}{N^2} (\mu_i \mu_j)^2 \ophi^2 \oepsilon^2 
G^P_n(\alpha_{si},\alpha_{sj})\nonumber \\
&& \times f^G( \Delta_{ij},\Delta_{ik}, \Delta_{jk}) \;\; .
\end{eqnarray} 
The function $f^G( \Delta_{ij},\Delta_{ik}, \Delta_{jk})$ is given in the
appendix. The contribution to the free energy is according to 
Eq.~(\ref{Fseries}) and Eq.~(\ref{FtermsinOphi2}) given by
\begin{eqnarray}
&& - k T \frac{1}{6} {\sum}^\prime 
3 \left< f^{(2)}_{ij} f^{(0)}_{jk} f^{(0)}_{ki} \right>_0 \nonumber \\ 
&&\qquad = - N k T \ophi^2 \oepsilon^2 \frac{1 + 6 \ln 2}{4} 
\widetilde{G}_{\text{poly},2} (\oalpha_s) 
\label{FcontriG} \;\; .
\end{eqnarray}
The function
\begin{eqnarray}\label{Gpoly2tilde}
&& \widetilde{G}_{\text{poly},2} (\oalpha_s) = \frac{6}{1 + 6 \ln 2}\\ 
&& \qquad \times \frac{1}{N^3} {\sum}^\prime
(\mu_i \mu_j)^2 G^P_n(\alpha_{si},\alpha_{sj}) 
f^G( \Delta_{ij},\Delta_{ik}, \Delta_{jk}) \;\; , \nonumber
\end{eqnarray}
was defined in such a way, that it reduces to $G_2(\alpha_s)$ in the
monodisperse case. Finally introducing the diameter distribution function 
$P(\Delta)$ requires the replacement
\begin{equation}
\frac{1}{N^3} {\sum}^\prime \rightarrow 
\int P(\Delta_i) \,d \Delta_i P(\Delta_j) \,d \Delta_j 
P(\Delta_k) \,d \Delta_k \label{Pdintro} 
\end{equation}
in eqs.~(\ref{FcontriG},\ref{Gpoly2tilde}).

\subsubsection{Graph H}

The integrations to calculate 
$\left< f^{(1)}_{ij} f^{(1)}_{jk} f^{(0)}_{ki} \right>_0$ are performed as
follows (compare to appendix B.8 in paper~I): After doing the trivial
integrations (concerning the possible configurations of the particles except 
$i$, $j$, and $k$, and the center of mass of the cluster), the possible 
orientations of the cluster are integrated out. Then, the integrations over the
orientations of ${\bf m}_i$, ${\bf m}_j$, and ${\bf m}_k$ are performed.
One arrives at
\begin{eqnarray}
&& \left< f_{ij}^{(1)} f_{ik}^{(1)} f_{jk}^{(0)} \right>_0 = 
\mu_i^2 \mu_j \mu_k \oepsilon^2 \frac{4 \pi^2}{15 V^2} 
K^P(\alpha_{si}, \alpha_{sj}, \alpha_{sk}) 
\nonumber \\
&& \qquad \times \int 
\frac{\oD^6}{r_{ij} r_{ik}} \left( 3 \cos^2 \vartheta_{jk} - 1 \right)
e^{-v_{ij}^{HC}} e^{-v_{ik}^{HC}} f_{jk}^{(0)} 
\nonumber \\
&& \qquad \qquad \times \dd r_{ij} 
\dd r_{ik} \sin \vartheta_{jk} \dd \vartheta_{jk} \;\; . \qquad\qquad
\label{Hresult1}
\end{eqnarray}
This is up to a factor $N^2 Z_0/2$ a polydisperse generalization of 
Eq.~(B21) of paper~I. The remaining integrations concern the distances $r_{ij}$
and $r_{ik}$, and the angle $\vartheta_{jk}$ between ${\bf r}_{ij}$ and
${\bf r}_{ik}$. 
The function $K^P$ is defined by
\begin{eqnarray}
&& K^P(\alpha_{si}, \alpha_{sj}, \alpha_{sk}) 
= \frac{3}{8} \frac{(4 \pi V)^3}{z_i z_j z_k}
\nonumber \\ && \qquad \times
\int_{-1}^1 \int_{-1}^1 \int_{-1}^1
e^{\left( \alpha_{si} u_i + \alpha_{sj} u_j +\alpha_{sk} u_k \right)} 
\\ && \qquad \qquad \times
\left( 3 + u^2_i \right) u_j u_k \, \dd u_i \dd u_j \dd u_k 
\nonumber
\label{Kdef}
\end{eqnarray}
[compare to Eq.~(B22) of paper~I]. An expression for the monodisperse
counterpart $K(\alpha_s)$ was given in appendix A of paper~I. Here we express $K^P$
via Langevin functions: 
\begin{eqnarray}
&& K^P(\alpha_{si}, \alpha_{sj}, \alpha_{sk}) =  \nonumber \\ 
&& \qquad 3 {\cal L}(\alpha_{sj}) {\cal L}(\alpha_{sk}) 
\left[ 3 + {\cal L}^\prime(\alpha_{si}) + {\cal L}(\alpha_{si})^2 \right]\;\; .
\label{Kresult}
\end{eqnarray}
Note that $r_{ij} > D_{ij}$ and $r_{ik} > D_{ik}$ is required, otherwise the 
integral (\ref{Hresult1}) vanishes. The requirement that particles $j$ and $k$ 
have to overlap imposes the additional restrictions 
$|r_{ij} - r_{ik}| < D_{jk}$ and  
\begin{equation}
\vartheta_{jk} < \vartheta^{\text{max}}_{jk} = \arccos 
\frac{r_{ij}^2 + r_{ik}^2 -D_{jk}^2}{2 r_{ij} r_{ik}} \;\; .
\end{equation}
After performing the remaining integrations within these limits the result is 
\begin{eqnarray}
&&\left< f_{ij}^{(1)} f_{ik}^{(1)} f_{jk}^{(0)} \right>_0 = 
\mu_i^2 \mu_j \mu_k \ophi^2 \oepsilon^2 \frac{48}{5 N^2} \nonumber \\
&& \qquad \times 
K^P(\alpha_{si}, \alpha_{sj}, \alpha_{sk}) 
f^K(\Delta_{ij}, \Delta_{ik}, \Delta_{jk}) \;\; . \label{Hresult2}
\end{eqnarray}
The function $f^K(\Delta_{ij}, \Delta_{ik}, \Delta_{jk})$ is given in the 
appendix. 
The contribution to the free energy is
\begin{equation}
- k T \frac{1}{6} {\sum}^\prime 
3 \left< f^{(1)}_{ij} f^{(1)}_{jk} f^{(0)}_{ki} \right>_0 = 
N k T \ophi^2 \oepsilon^2 K_{\text{poly}} (\oalpha_s) \;\; ,
\label{FcontriH}
\end{equation}
with
\begin{eqnarray}
&& K_{\text{poly}} (\oalpha_s) = -\frac{24}{5} \label{Kpolydef}  
\frac{1}{N^3} {\sum}^\prime \mu_i^2 \mu_j \mu_k 
\\ && \qquad \times
K^P(\alpha_{si},\alpha_{sj},\alpha_{sk}) 
f^K( \Delta_{ij},\Delta_{ik}, \Delta_{jk}) 
\;\; . \nonumber
\end{eqnarray}

\subsubsection{The magnetization contribution}

Inserting Eq.~(\ref{FcontriG}) and Eq.~(\ref{FcontriH}) in Eq.~(\ref{Fseries})
and calculating the equilibrium magnetization results in an additional
$O(\ophi^2 \oepsilon^2)$--term in Eq.~(\ref{MofalphasOphi}):
\begin{eqnarray}
\frac{N \om}{\mu_0 V} \ophi^2 \oepsilon^2 
\left[ \frac{1 + 6 \ln 2}{4} 
\widetilde{G}_{\text{poly},2}^\prime (\oalpha_s) - 
K_{\text{poly}}^\prime (\oalpha_s) \right] \;\; .
\end{eqnarray}
The expansion and iteration procedure to switch from $\oalpha_s$ to $\oalpha$
is identical to the monodisperse case in paper~I. The full expression for the 
magnetization containing all calculated terms reads
\begin{eqnarray}\label{Mfinal}
M &=& 
\frac{N \om}{\mu_0 V}  \Biggl[ {\cal L}_{\text{poly}}(\oalpha) + 
8 \ophi \oepsilon {\cal L}_{\text{poly}} (\oalpha) 
{\cal L}_{\text{poly}}^\prime (\oalpha) \nonumber \\
&& \qquad + \sum_{n=2}^\infty
\ophi \oepsilon^n G_{\text{poly},n}^\prime (\oalpha) \nonumber \\
&& \qquad + 64 \ophi^2 \oepsilon^2 {\cal L}_{\text{poly}} (\oalpha) 
{\cal L}_{\text{poly}}^\prime (\oalpha)^2\\
&& \qquad + 32 \ophi^2 \oepsilon^2 {\cal L}_{\text{poly}} (\oalpha)^2 
{\cal L}^{\prime\prime}_{\text{poly}} (\oalpha) \nonumber \\
&& \qquad + \ophi^2 \oepsilon^2 \frac{1 + 6 \ln 2}{4} 
\widetilde{G}_{\text{poly},2}^\prime (\oalpha)  - 
\ophi^2 \oepsilon^2 K_{\text{poly}}^\prime (\oalpha) \Biggr]
\;\; .\nonumber
\end{eqnarray}

\section{Selected results for the magnetization}
\label{SEC4}

\begin{figure}
\includegraphics[width=16cm,angle=0]{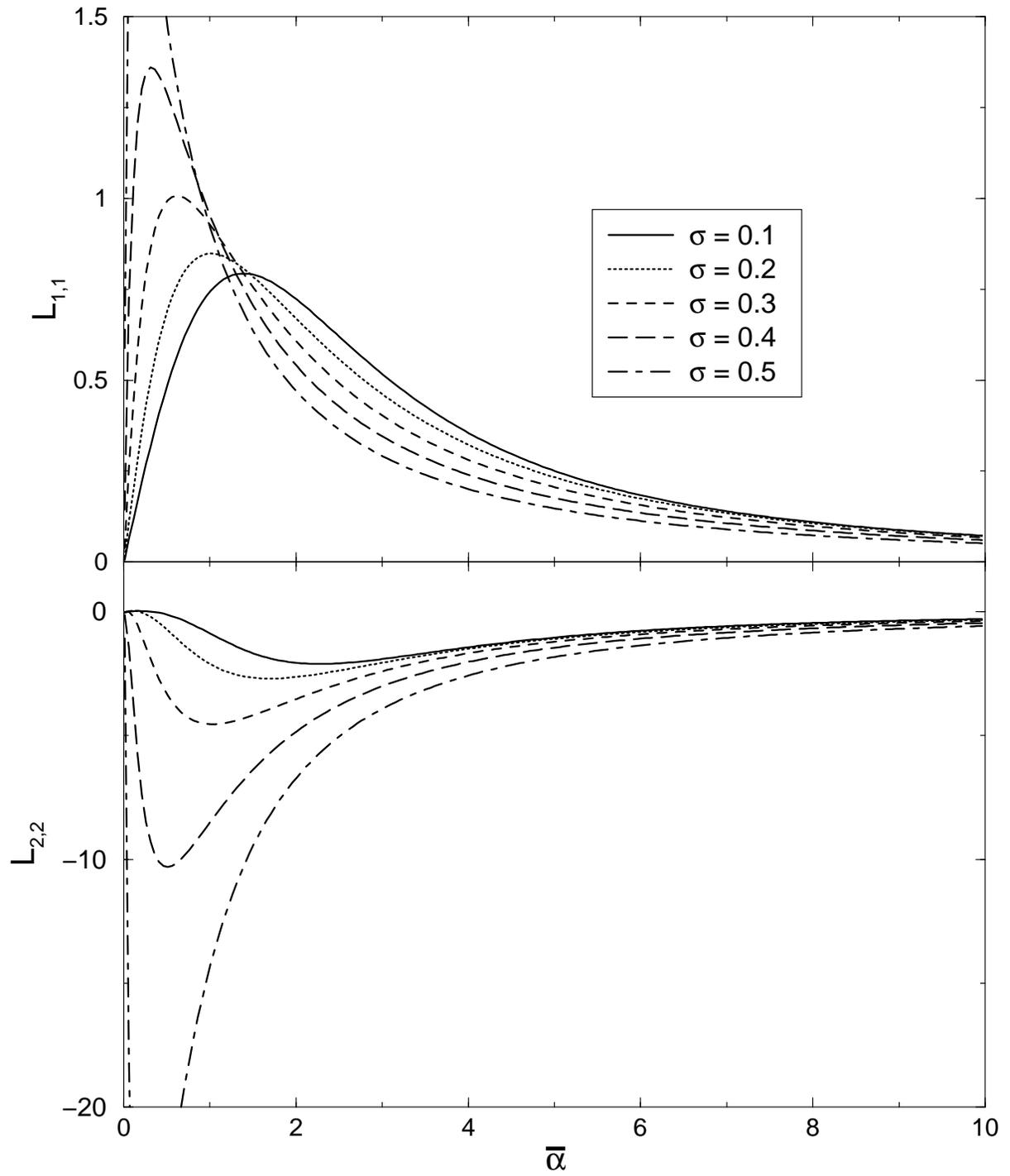}
\caption[]
{The contributions $L_{1,1}$ and $L_{2,2}$ as functions of $\oalpha$ for
lognormal distributions with different widths $\sigma$. Here
$\left< \Delta^3 \right>_P = 1$ and $\mu(\Delta) = \Delta^3$.}
\label{L11L22fig} 
\end{figure}

\begin{figure}
\includegraphics[width=16cm,angle=0]{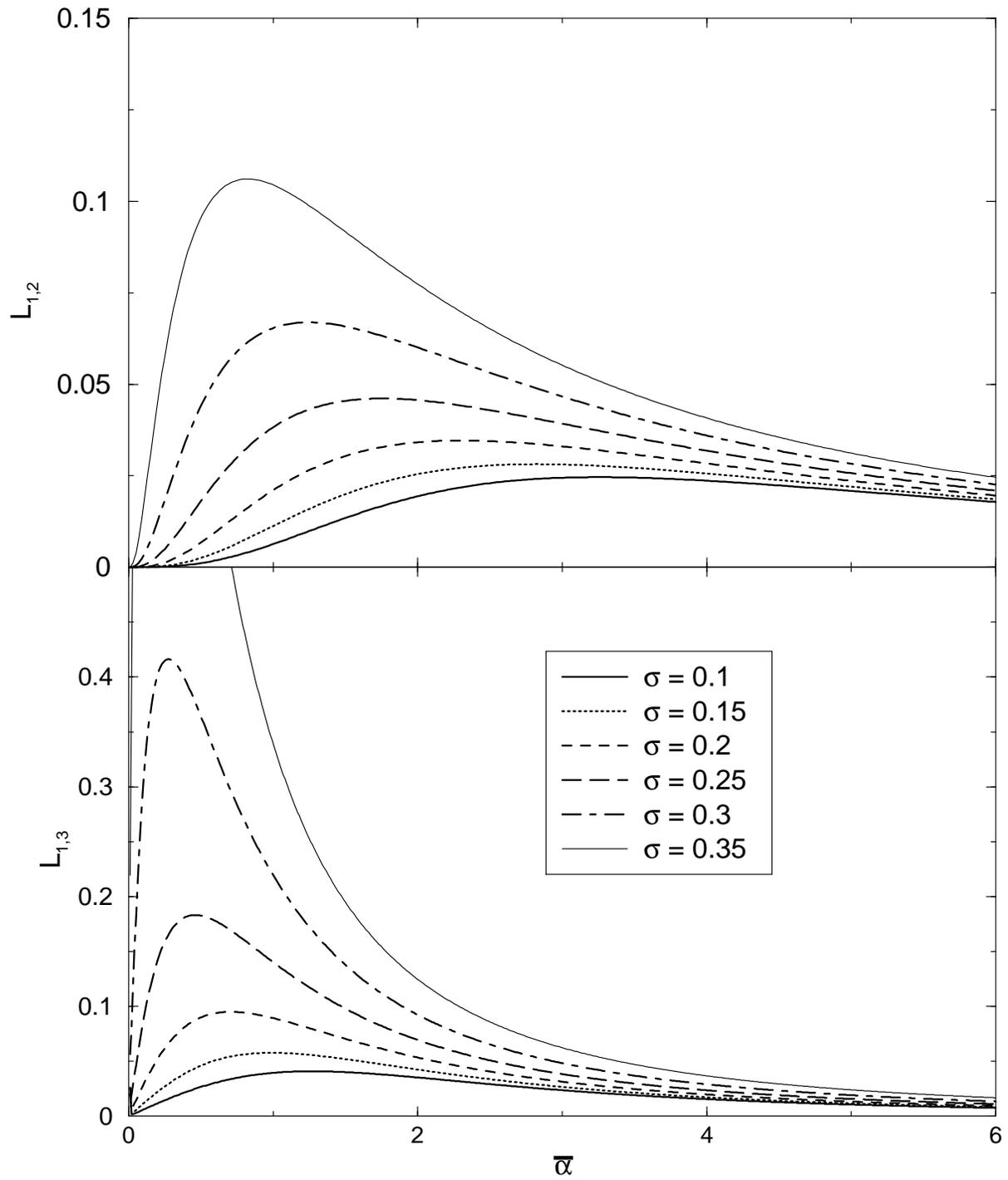}
\caption[]
{The contributions $L_{1,2}$ and $L_{1,3}$ as a function of $\oalpha$ for
lognormal distributions with different widths $\sigma$. $\Delta$ and 
$\mu(\Delta)$ as in Fig.~\ref{L11L22fig}.}
\label{L13fig}
\end{figure}

The figures \ref{L11L22fig} and \ref{L13fig} showing 
separate contributions to $M(\oalpha)$ in polydisperse systems were obtained 
by using lognormal distributions
\begin{equation} \label{logndist}
P(D) = \frac{1}{\sqrt{2 \pi} \sigma D_0 } e^{-\sigma^2/2}
e^{-\ln^2(D/D_0)/(2 \sigma^2)}
\end{equation}
for the particle diameters as a representative and often used example for size 
distibutions of model polydisperse systems. Using similar distributions like, 
e.~g., \ Gamma distributions, however, would not qualitatively modify the results. 
In Sec.~\ref{SEC4a} we use also an experimentally determined size distribution.
Here, the quantity $\oD$ was taken to 
be defined via the mean volume  
(\ref{defDbar}), i.e., $\oD^3 = \left< D^3 \right>_P =
D_0^3 e^{15 \sigma^2/2}$ so that $\ophi$ is 
the volume fraction 
$\phi$. The magnetic moments of the particles were taken to scale with their
volumes, $m \sim D^3$, allowing to set $\mu(\Delta) = \Delta^3$ such that
$\om = \left< m \right>_P$.

Figures \ref{L11L22fig} and \ref{L13fig} show the contributions
\begin{mathletters}
\begin{eqnarray}
L_{1,1} (\oalpha) &=& 8 {\cal L}_{\text{poly}} (\oalpha) 
{\cal L}_{\text{poly}}^\prime (\oalpha) \;\; ,\\
L_{2,2} (\oalpha) &=&  64 {\cal L}_{\text{poly}} (\oalpha) 
{\cal L}_{\text{poly}}^\prime (\oalpha)^2 \nonumber \\
&& + 32 {\cal L}_{\text{poly}} (\oalpha)^2 
{\cal L}^{\prime\prime}_{\text{poly}} (\oalpha)  \\
&& + \frac{1 + 6 \ln 2}{4} 
\widetilde{G}_{\text{poly},2}^\prime (\oalpha)  - 
K_{\text{poly}}^\prime (\oalpha) \;\; , \nonumber \\
L_{1,2} (\oalpha) &=& G_{\text{poly},2}^\prime (\oalpha)\;\; ,\\
L_{1,3} (\oalpha) &=& G_{\text{poly},3}^\prime (\oalpha)
\end{eqnarray}
\end{mathletters}
to $M(\oalpha)$ (\ref{Mfinal})
for different values of the width $\sigma$ of the distribution (\ref{logndist}).
The contributions of higher-order terms increase with growing $\sigma$. 
This is so because they depend
on higher moments of the distribution $P(\Delta)$ that grow with 
the width of the distribution, even if the third moment 
$\left< \Delta^3 \right>_P$ is kept fixed. The shift of the maxima
of the curves to the right has a similar reason: Bigger particles, that react to smaller
fields get more and more important when the width of the distribution
grows.

The assumption $\mu(\Delta) = \Delta^3$ that the magnetic moments in our 
ferrofluid model scale $\sim D^3$ with the total volume 
of its hard sphere constituents is somewhat too simple for particles in real
ferrofluids for two reasons: First, in the common case of steric stabilization 
by polymers surfactants providing the repulsion the surfactant layer of about 
1 -- 3 nm does not contribute to the magnetic moment. Second, an outer layer of
the magnetic material might be magnetically dead so that it does not contribute
to the magnetic moment either. For magnetite particles a dead layer depth 
of $\approx$ 0.8 nm has been reported \cite{SISI87}. To account for the sum of 
these
two effects we have introduced in our calculation for $M$ an effective magnetic 
diameter $D_{\text{mag}}$ via the relation $D = D_{\text{mag}} + 5.6 \text{ nm}$.
It ascribes to every particle that has a magnetically effective core of diameter 
$D_{\text{mag}}$ a hard sphere with a magnetically inert layer of depth 2.8 nm.
The magnetic moment of each particle was then taken to scale with is magnetically
effective volume, i.e., $\mu = (D_{\text{mag}}/\oD)^3$.

\begin{figure}
\includegraphics[width=12cm,angle=270]{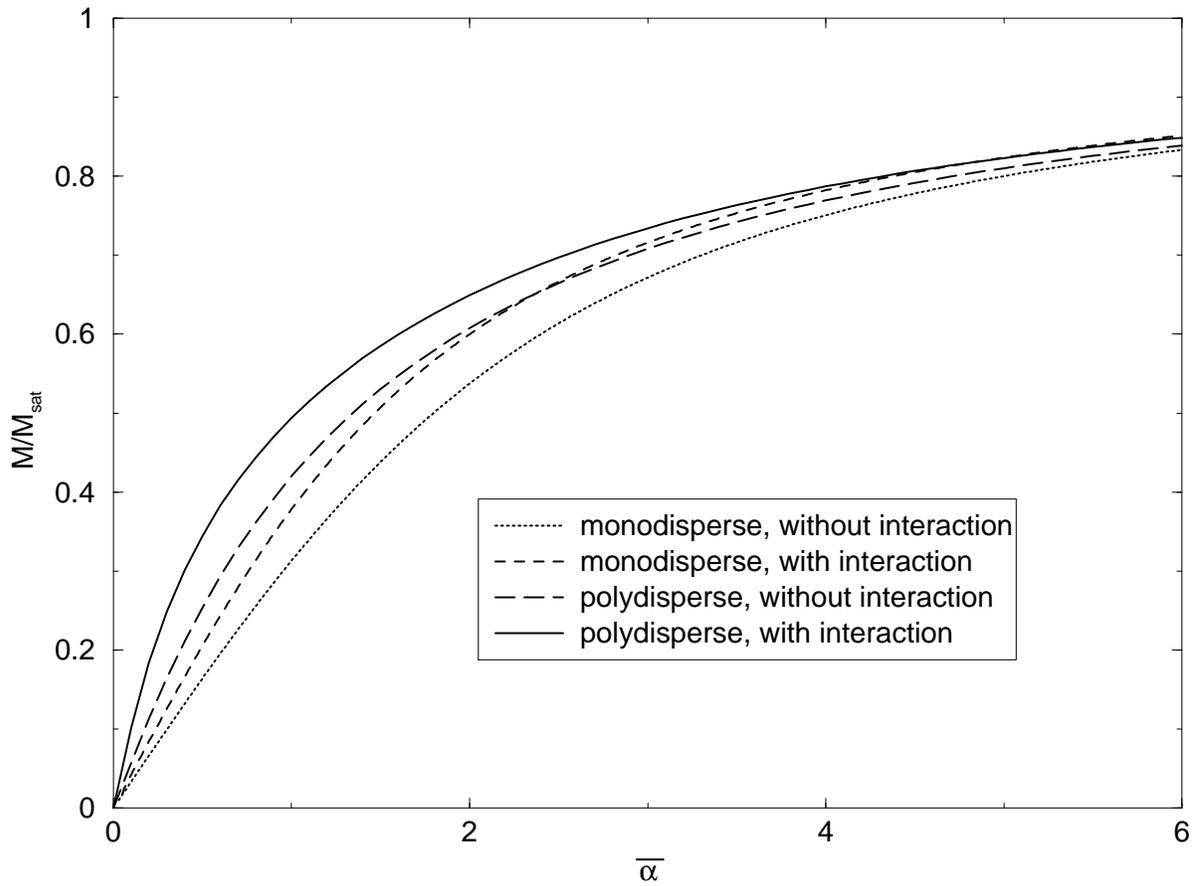}
\caption[]
{Equilibrium magnetization of a polydisperse ferrofluid 
($\left< D_{\text{mag}}^3 \right>_P^{1/3}$ = 10 nm, $\sigma = 0.25$, 
$\mu = (D_{\text{mag}}/\oD)^3$)
and a comparable monodisperse ferrofluid, both with and without 
taking into account the particle interaction. See text for further details.}
\label{magcomp}
\end{figure}

The full line in Fig.~\ref{magcomp} shows the reduced equilibrium magnetization
of an interacting polydisperse ferrofluid
as function of $\oalpha$. Here each particle has a total magnetic inactive layer of 
2.8 nm and the diameters $D_{\text{mag}}$ of the magnetic 
cores are lognormally distributed with $\sigma = 0.25$
and $\left< D_{\text{mag}}^3 \right>_P^{1/3}$ = 10 nm. The latter is 
taken as the reference diameter $\oD$ in the calculation.
The particle density is chosen in 
such a way that $\ophi = 0.05$. Note that here $\ophi = \phi_{\text{mag}}$,
the volume fraction of the magnetically active material. The magnetic
moment $\om$ for $D_{\text{mag}} =$ 10 nm is chosen such that $\oepsilon = 2$.
The magnetization of the core is then about 550 kA/m at room temperature which
is a little bit larger than that of magnetite. 

Expansion terms of order 
$\ophi \oepsilon^n$ are taken into account up to order $n$=5 for the 
magnetization
curves in Fig.~\ref{magcomp}. Higher $O(\ophi \oepsilon^n)$--terms have only a 
small effect on the magnetization. Here $\oepsilon$ is by
definition of $\om$ and $\oD$ a typical interaction energy divided by $kT$ 
for particles at a distance of $\oD$ = 10 nm. But with the two additional dead 
layers
of total size 5.6 nm in between the particles of our model ferrofluid the real 
typical dipolar energies at contact are smaller. 

The magnetization is compared to that of a
polydisperse fluid without particle--particle interaction (long dashed line), 
and to a 
monodisperse ferrofluid, both with and without taking into account the 
particle--particle interaction (short dashed and dotted line, respectively). 
The monodisperse system consists of particles with $D_{\text{mag}} =$ 10 nm
with the same nonmagnetic layer thickness, bulk magnetization, and 
$\phi_{\text{mag}}$ as before. 

One sees that 
taking into account polydispersity or particle interaction alone strengthens
the magnetization and especially the initial susceptibility. Both effects are 
comparable for the given parameters. Together, they result in an even higher
equilibrium magnetization. 

\begin{figure}
\includegraphics[width=13cm,angle=0]{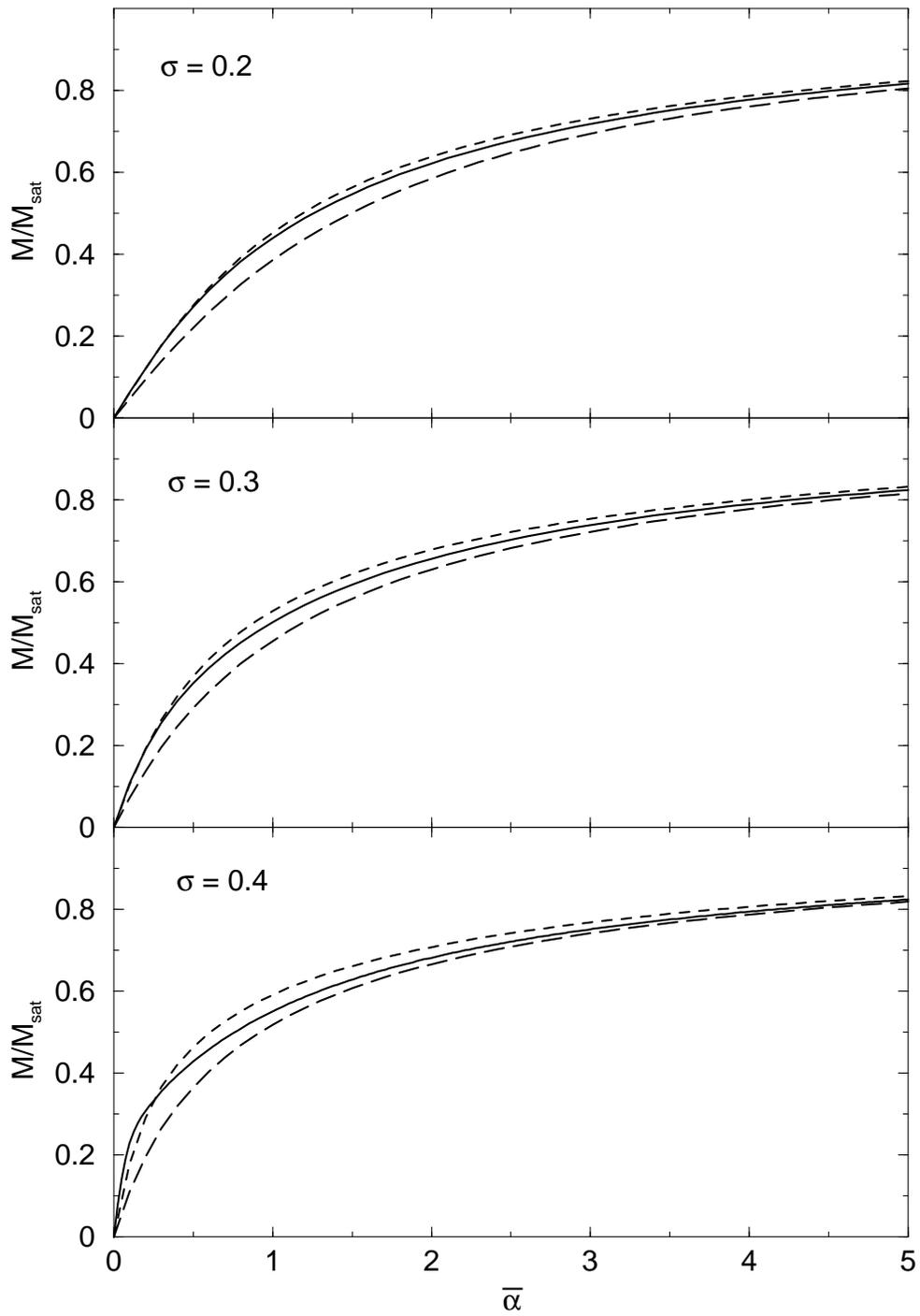}
\caption[] 
{Equilibrium magnetization as a function of $\oalpha$ for lognormal 
distributions with $\left< D_{\text{mag}}^3 \right>_P^{1/3}$ = 8 nm and different 
widths $\sigma$. Long dashed: without particle interaction, dashed: 
particle interaction only in $O(\ophi \oepsilon)$, solid lines: all calculated
terms.}
\label{magofsigma}
\end{figure}

Figure \ref{magofsigma} shows magnetization curves for magnetite--based
ferrofluids with a bulk magnetization of 480 kA/m for distributions of different
widths. $D_{\text{mag}}$ is taken 
to be lognormally distributed 
with $\left< D_{\text{mag}}^3 \right>_P = (8 \text{ nm})^3$ and $\sigma =$ 0.2,
0.3, and 0.4. The volume fraction
of the magnetic material is $\phi_{\text{mag}} = 0.1$. The particles are again 
assumed to carry a nonmagnetic layer of 2.8 nm thickness. 

The increase in $\sigma$ causes an increase of the initial
susceptibility already in the noninteracting case (long dashed lines). 
Including the  $O(\ophi \oepsilon)$--terms (short dashed lines) has a positive 
effect on the magnetization. The relative increase is maximal for small 
$\oalpha$. The magnetization decreases again at higher $\oalpha$, if higher 
order terms are taken into account (solid lines). For the considered 
ferrofluids the $O(\ophi^2 \oepsilon^2)$--term that is negative for higher
$\oalpha$ (see Fig.~\ref{L11L22fig}) is almost solely responsible for this 
decrease. The positive contributions from the higher 
$O(\ophi \oepsilon^n)$--terms ($n \geq 2$) are again negligible, except for 
$\sigma = 0.4$ and small $\oalpha$, 
where they cause a further increase of the initial susceptibility. 
For small $\oalpha$ the $O(\ophi^2 \oepsilon^2)$--term also has a positive 
effect, but this effect is too small to be visible. The plots show again,
that the influence of higher order terms is larger for broad distributions. 

\section{Comparison to experiments}
\label{SEC4a}

\begin{figure}
\includegraphics[width=15cm,angle=270]{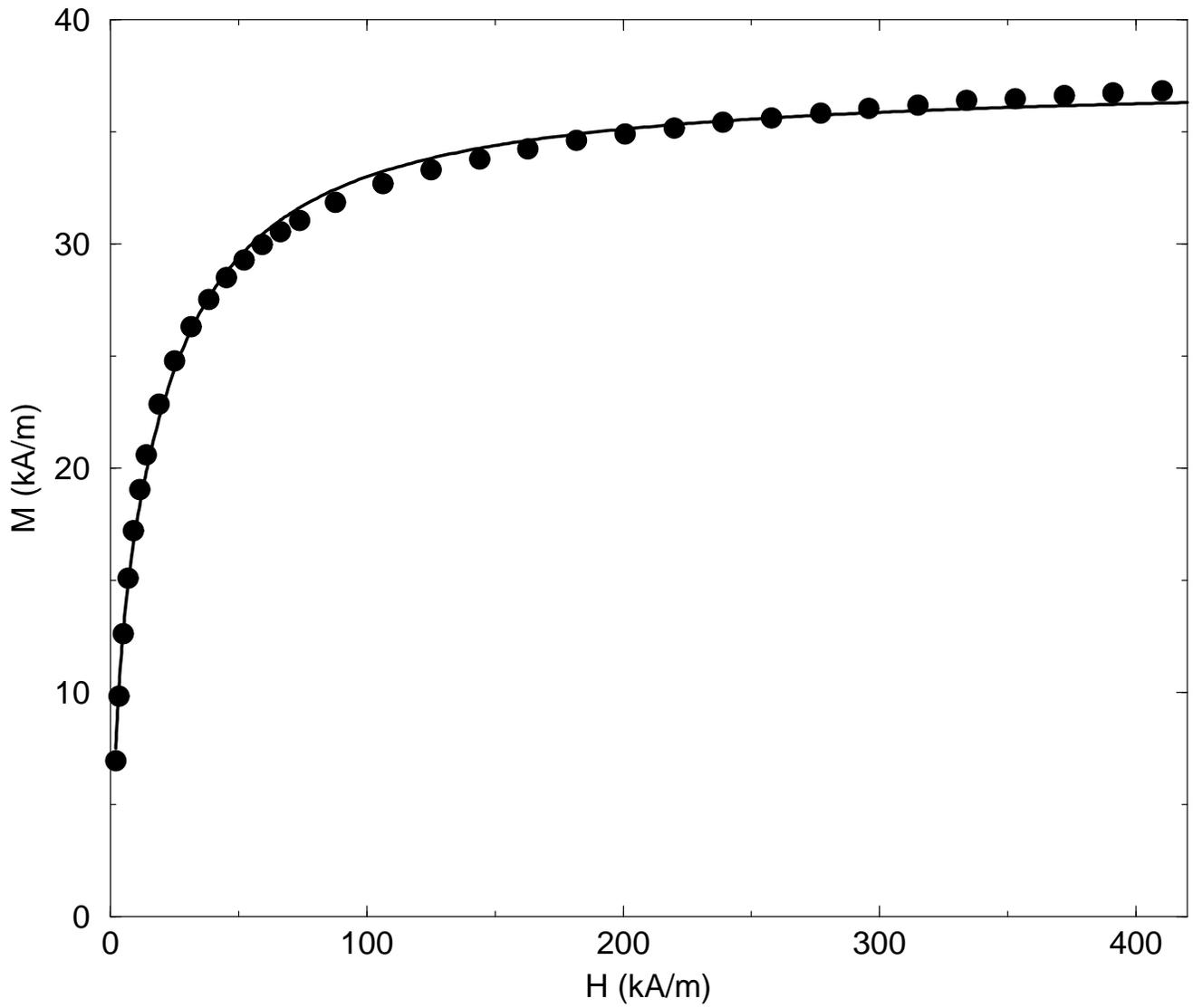}
\caption[] 
{Equilibrium magnetization of the ferrofluid EMC 905. Dots: experiment,
solid line: the theoretical magnetization curve assuming lognormally distributed
$D_{\text{mag}}$, c.f. text.}
\label{magEMC}
\end{figure}

We compared our theorerical predictions for the 
magnetization curves with experimental results of
two different magnetite--based ferrofluids. S. Odenbach (ZARM, Bremen) provided 
data on the equilibrium magnetization of
the ferrofluid EMC~905 produced by FerroTec. We fitted our theoretical result 
(\ref{Mfinal}) taking into account the terms in $O(\phi \epsilon^n)$ 
up to $n = 5$ to the data assuming lognormally distributed $D_{\text{mag}}$
and a nonmagnetic layer of depth 2.8 nm. The bulk magnetization of magnetite
was taken to be 480 kA/m. The result is shown in Fig.~\ref{magEMC}.
According to the fit, the saturation magnetization of the ferrofluid is
$M_{\text{sat}} = 37.4$ kA/m. The parameters defining the distribution turn out
to be $D_0 = 8.3$ nm and $\sigma = 0.28$. There are small differences between
the data and the fit curve that are in our opinion due to deviations of the real
diameter distribution curve from the idealized lognormal form. 

\begin{figure}
\includegraphics[width=12.5cm,angle=0]{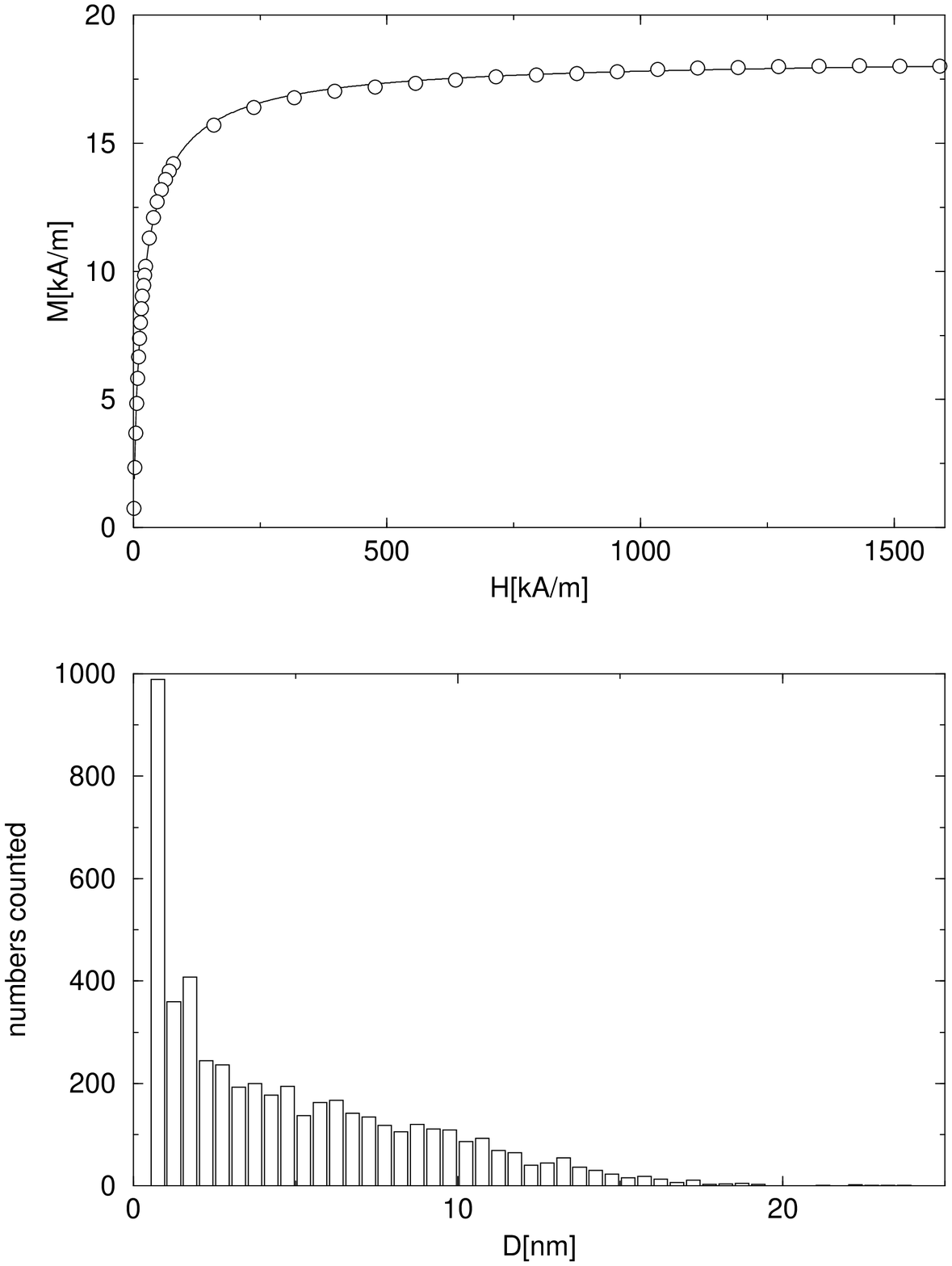}
\caption[] 
{Top: equilibrium magnetization of the ferrofluid APG 933. 
Circles: experiment, solid line: the theoretical magnetization curve. Bottom: 
diameter distribution according to TEM measurements.}
\label{magAPG}
\end{figure}

J. Embs (Universit\"at des Saarlandes, Saarbr\"ucken) measured the equilibrium
magnetization curve of the ferrofluid APG~933 of FerroTec. In addition he
determined the diameter distribution of its particles by transmission electron
microscopy (TEM) which was then used in our theoretical analysis. Diameters found 
in TEM measurements are those of the magnetite particles. We assumed the 
magnetically effective 
diameters $D_{\text{mag}}$ to be $2 \times 0.8$ nm $= 1.6$ nm smaller and to be 
zero for particles smaller than 1.6 nm.  The hard core diameters $D$ were taken
to be $2 \times 2$ nm $= 4$ nm larger than the diameters obtained from the TEM
measurements. As above, we took into account terms 
up to $O(\phi \epsilon^5)$ and set the bulk magnetization of magnetite to 
480 kA/m. Fig.~\ref{magAPG} shows the TEM data and 
the experimental magnetization curve together with the results of our theory. 
Both agree very well.

\section{Conclusion}
\label{SEC5}

In this paper we used the technique of cluster expansion to derive 
an approximation to the equilibrium magnetization for the system of dipolar
hard spheres in a magnetic field with diameter and/or magnetic moment 
dispersion as a model system for a polydisperse ferrofluid. The calculation
results in an expression for the magnetization $M$ in form of a twofold 
series expansion in the parameters $\ophi$, closely related to the volume 
fraction $\phi$, and $\oepsilon$, a coupling
parameter measuring the strength of the dipolar interaction:
\begin{equation}
M = \frac{N \om}{\mu_0 V}  
\sum_{m=0}^\infty \sum_{n=0}^\infty \ophi^m \oepsilon^n
L_{m,n} (\oalpha) \;\; .
\end{equation}
$\ophi$, $\oepsilon$, and the dimensionless magnetic field $\oalpha$ are 
defined for some typical values $\oD$ and $\om$ for the hard sphere diameters
and magnetic moments respectively. $\om$ can be chosen in such a way that
the prefactor $N \om/\mu_0 V$ reduces to the saturation magnetization of the
system. We gave expressions for $L_{1,n}$ ($n \leq 5$) and $L_{2,2}$. Lower
orders vanish, except for $L_{0,0}$ reducing to the Langevin function in the
monodisperse case. The calculated $L_{m,n}$ can be written as multiple sums
over all particles whose addends are analytical expressions.

The influence of particle--particle interaction grows with increasing width
of the considered diameter distribution.
Taking into account only the $L_{1,1}$--term results in an increase of the 
magnetization relative to the non interacting system, 
whereas the $L_{2,2}$--term leads again to somewhat smaller values at higher
$\oalpha$. Only at very small $\oalpha$ its contribution is positive. 
The $L_{1,n}$--terms have little effect for realistic, 
magnetite--based ferrofluids, except for broad distributions, where they 
increase the initial magnetization.

\acknowledgments

We would like to thank Jan Embs and Stefan Odenbach for providing
the experimental data discussed in Sec.~\ref{SEC4a}.
This work was supported by the Deutsche Forschungsgemeinschaft (SFB 277).

\appendix

\section{The functions $G_n^P, f^G, f^K$}
\label{appa}


The functions $G_n^P(x_1,x_2)$ are symmetric in their
arguments and have the form 
\begin{eqnarray}
&& G_n^P (x_1,x_2 ) = 
G^{P(0)}_{n}( 1/x_1 , 1/x_2) \nonumber \\ 
&& \qquad + \coth (x_1) G^{P(1)}_{n}(1/x_1, 1/x_2) 
\\ 
&& \qquad + \coth (x_2 ) G^{P(1)}_{n}(1/x_2, 1/x_1) 
\qquad \nonumber \\ && \qquad +
\coth (x_1 ) \coth (x_2 )  
G^{P(2)}_{n}(1/x_1 , 1/x_2 ) \;\; .\nonumber 
\end{eqnarray}
The $G^{P(i)}_{n}$ are polynomials and read for $n \leq 5$
\begin{mathletters}
\begin{eqnarray}
G_2^{P(0)}(y_1,y_2) &=&
\frac{8}{5}  
+\frac{4}{5} y_1^{2} 
+\frac{4}{5}  y_2^{2}
+\frac{12}{5} y_1^{2} y_2^{2} \;\; ,
\nonumber \\
G_2^{P(1)}(y_1,y_2) &=&
-\frac{4}{5} y_1 
-\frac{12}{5} y_1 y_2^{2} \;\; ,
\\
G_2^{P(2)}(y_1,y_2) &=&
\frac{12}{5}  \;\; .
\nonumber
\end{eqnarray}
\begin{eqnarray}
G_3^{P(0)}(y_1,y_2) &=&
-\frac{4}{35} y_1 y_2
-\frac{24}{35} y_1^{3} y_2 \nonumber\\ &&
-\frac{24}{35} y_1 y_2^{3}
-\frac{12}{7} y_1^{3} y_2^{3} \;\; ,
\nonumber \\
G_3^{P(1)}(y_1,y_2) &=&
-\frac{4}{35}  y_2
+\frac{24}{35} y_1^{2} y_2 \\ && \nonumber
+\frac{4}{35}  y_2^{3}
+\frac{12}{7} y_1^{2} y_2^{3} \;\; ,
\\
G_3^{P(2)}(y_1,y_2) &=&
\frac{16}{105}  
-\frac{4}{35} y_1^{2} 
-\frac{4}{35}  y_2^{2}
-\frac{12}{7} y_1^{2} y_2^{2} \;\; .
\nonumber
\end{eqnarray}
\begin{eqnarray}
G_4^{P(0)}(y_1,y_2) &=&
\frac{8}{105}  
+\frac{4}{35} y_1^{2} 
+\frac{4}{35} y_1^{4} 
+\frac{4}{35}  y_2^{2}\nonumber \\ &&
+\frac{12}{5} y_1^{2} y_2^{2} 
+\frac{36}{7} y_1^{4} y_2^{2}
+\frac{4}{35}  y_2^{4} \nonumber \\ &&
+\frac{36}{7} y_1^{2} y_2^{4}
+12 y_1^{4} y_2^{4}\;\; ,
\nonumber \\
G_4^{P(1)}(y_1,y_2) &=&
-\frac{8}{105} y_1 
-\frac{4}{35} y_1^{3} 
-\frac{24}{35} y_1 y_2^{2}\\ &&
-\frac{36}{7} y_1^{3} y_2^{2}
-\frac{8}{7} y_1 y_2^{4}
-12 y_1^{3} y_2^{4} \;\; ,\nonumber
\\
G_4^{P(2)}(y_1,y_2) &=&
\frac{32}{105} y_1 y_2
+\frac{8}{7} y_1^{3} y_2
+\frac{8}{7} y_1 y_2^{3}
+12 y_1^{3} y_2^{3} \;\; .
\nonumber
\end{eqnarray}
\begin{eqnarray}
G_5^{P(0)}(y_1,y_2) &=&
\frac{12}{385} y_1 y_2
-\frac{104}{385} y_1^{3} y_2
-\frac{60}{77} y_1^{5} y_2\nonumber \\ &&
-\frac{104}{385} y_1 y_2^{3}
-\frac{732}{77} y_1^{3} y_2^{3}
-\frac{240}{11} y_1^{5} y_2^{3}\nonumber \\ &&
-\frac{60}{77} y_1 y_2^{5}
-\frac{240}{11} y_1^{3} y_2^{5}
-\frac{540}{11} y_1^{5} y_2^{5} \;\; ,
\nonumber \\
G_5^{P(1)}(y_1,y_2) &=&
-\frac{4}{231}  y_2
+\frac{4}{385} y_1^{2} y_2
+\frac{60}{77} y_1^{4} y_2 \nonumber \\ &&
+\frac{4}{385}  y_2^{3}
+\frac{172}{77} y_1^{2} y_2^{3}
+\frac{240}{11} y_1^{4} y_2^{3}  \\ &&
+\frac{4}{77}  y_2^{5}
+\frac{60}{11} y_1^{2} y_2^{5}
+\frac{540}{11} y_1^{4} y_2^{5} \;\; ,\nonumber 
\\
G_5^{P(2)}(y_1,y_2) &=&
\frac{16}{1155}  
+\frac{8}{1155} y_1^{2} 
-\frac{4}{77} y_1^{4} \nonumber \\ &&
+\frac{8}{1155}  y_2^{2}
-\frac{32}{77} y_1^{2} y_2^{2}
-\frac{60}{11} y_1^{4} y_2^{2}\nonumber \\ &&
-\frac{4}{77}  y_2^{4} 
-\frac{60}{11} y_1^{2} y_2^{4}
-\frac{540}{11} y_1^{4} y_2^{4} \;\; .
\nonumber
\end{eqnarray}
\end{mathletters}

It is
\begin{eqnarray}
&&f^G( \Delta_{ij},\Delta_{ik}, \Delta_{jk}) = 
\ln \left( \frac{\Delta_{ik} + \Delta_{jk}}{\Delta_{ij}} \right)
- \frac{ 3( \Delta_{ik}^2 - \Delta_{jk}^2)^2}
{4 \Delta_{ij}^4} 
\nonumber \\ && \qquad 
+ \frac{ 8 ( \Delta_{ik}^3 + \Delta_{jk}^3)}
{3 \Delta^3_{ij} }
-\frac{3( \Delta_{ik}^2 + \Delta_{jk}^2)}{\Delta_{ij}^2} 
\\ && \qquad 
+ \frac{ 13 \Delta_{ik}^2 + 14 \Delta_{ik} \Delta_{jk} + 13 \Delta_{jk}^2}
{12 ( \Delta_{ik} + \Delta_{jk})^2} \;\; , \nonumber
\end{eqnarray}
and
\begin{eqnarray}
&& f^K(\Delta_{ij}, \Delta_{ik}, \Delta_{jk}) = -\frac{2}{9} 
- \frac{5}{36} \frac{\Delta_{ij}^6 + \Delta_{ik}^6 + 
\Delta_{jk}^6}{\Delta_{ij}^3 \Delta_{ik}^3} 
\nonumber \\ && \qquad
-\frac{\Delta_{jk}^2}{4 \Delta_{ij} \Delta_{ik}} - 
\frac{2 \Delta_{jk}^3}{9 \Delta_{ij}^3} 
- \frac{2 \Delta_{jk}^3}{9 \Delta_{ik}^3} \\
&&\qquad + \frac{(\Delta_{ij}^4 + \Delta_{ik}^4 + \Delta_{jk}^4)
(\Delta_{ij}^2 + \Delta_{ik}^2 + \Delta_{jk}^2)}
{8 \Delta_{ij}^3 \Delta_{ik}^3} \;\; . \nonumber
\end{eqnarray}
When all diameters are equal, these functions reduce to
\begin{eqnarray}
f^G(\Delta,\Delta,\Delta) &=& \frac{1}{6} + \ln 2 \;\; ,\\
f^K(\Delta,\Delta,\Delta) &=& -\frac{5}{24} \;\; .
\end{eqnarray}

\clearpage

\clearpage
\begin{thebibliography}{\addcontentsline{toc}{section}{RERFERENCES}}

\bibitem{HL}
B.~Huke and M.~L\"ucke,
Phys.~Rev.~E~{\bf 62}, 6875 (2000).

\bibitem{R85}
R.~E.~Rosensweig,
{\it Ferrohydrodynamics},
(Cambridge University Press, Cambrigde, U.K., 1985).

\bibitem{W71} 
M.~S.~Wertheim,
J.~Chem.~Phys.~{\bf 55}, 4291 (1971).

\bibitem{AD73}
S.~A.~Adelman and J.~M.~Deutch,
J.~Chem.~Phys.~{\bf 59}, 3971 (1973).

\bibitem{IB74}
D.~Isbister and R.~J.~Bearman,
Molec.~Phys.~{\bf 28}, 1297 (1974).

\bibitem{HS78}
J.~S.~Hoye and G.~Stell, 
J.~Chem.~Phys.~{\bf 70}, 2894 (1978).

\bibitem{FHI79}
B.~Freasier, N.~Hamer, and D.~Isbister,
Molec.~Phys.~{\bf 38}, 1661 (1979).

\bibitem{RH81}
J.~D.~Ramshaw and N.~D.~Hamer,
J.~Chem.~Phys.~{\bf 75}, 3511 (1981).

\bibitem{CB86}
P.~T.~Cummings and L.~Blum,
J.~Chem.~Phys.~{\bf 85}, 6158 (1986).

\bibitem{FP84}
P.~H.~Fries and G.~N.~Patey,
J.~Chem.~Phys.~{\bf 82}, 429 (1984).

\bibitem{LL87}
P.~H.~Lee and B.~M.~Ladanyi,
J.~Chem.~Phys.~{\bf 87}, 4093 (1987).

\bibitem{LL89}
P.~H.~Lee and B.~M.~Ladanyi,
J.~Chem.~Phys.~{\bf 91}, 7063 (1989). 

\bibitem{KS02}
G.~Kronome, I.~Szalai, and J.~Liszi,
J.~Chem.~Phys.~{\bf 116}, 2067 (2002). 

\bibitem{K99}
V.~I.~Kalikmanov, 
Phys.~Rev.~E {\bf 59}, 4085 (1999).

\bibitem{SPMS90}
M.~I.~Shliomis, A.~F.~Pshenichnikov, K.~I.~Morozov, and I.~Yu.~Shurubor,
J.~Magn.~Magn.~Mater.~{\bf 85}, 40 (1990).

\bibitem{ML90}
K.~I.~Morozov and A.~V.~Lebedev,
J.~Magn.~Magn.~Mater.~{\bf 85}, 51 (1990).

\bibitem{BI92}
Y.~A.~Buyevich and A.~O.~Ivanov,
Physica A, {\bf 190}, 276 (1992).

\bibitem{PML96}
A.~F.~Pshenichnikov,  V.~V.~Mekhonoshin, and A.~V.~Lebedev,
J.~Magn.~Magn.~Mater.~{\bf 161}, 94 (1996).

\bibitem{IK01}
A.~O.~Ivanov and O.~B.~Kuznetsova,
Phys.~Rev.~E~{\bf 64}, 041405 (2001).

\bibitem{Ce82}
A.~O.~Cebers,
Magn.~Gidrodinamika {\bf 2}, 42 (1982).

\bibitem{BGW98}   
S.~Banerjee, R.~B.~Griffiths, and M.~Widom,
J.~Stat.~Phys.~{\bf 93}, 109 (1998).

\bibitem{SISI87}
T.~Sato, T.~Iijima, M.~Seki, and N.~Inagaki:
J.~Magn.~Magn.~Mater.~{\bf 65}, 252 (1987).

\end{thebibliography}
\end{document}